\documentclass[]{aastex631}
\usepackage{graphicx} 
\usepackage{natbib}
\usepackage{xspace}
\usepackage{amsmath}
\usepackage{hyperref}
\graphicspath{{./figures}}

\newcommand{\cyg}{Cyg\,X-3\xspace}
\newcommand{\xrism}{\textit{XRISM}\xspace}
\newcommand{\rsl}{\xrism/Resolve\xspace}

\newcommand{\chandra}{\textit{Chandra}\xspace}
\newcommand{\xspec}{\texttt{xspec}\xspace}
\newcommand{\isis}{\texttt{isis}\xspace}
\newcommand{\xstar}{\textsc{xstar}\xspace}
\newcommand{\spex}{SPEX\xspace}
\newcommand{\warmabs}{\texttt{warmabs}\xspace}
\newcommand{\photemis}{\texttt{photemis}\xspace}
\newcommand{\logxi}{\ensuremath{\log\xi}\xspace}
\newcommand{\kms}{\ensuremath{\mathrm{km}\,\mathrm{s}^{-1}}\xspace}

\begin{document}
\title{The \rsl view of the Fe\,K region of \cyg}

\author{XRISM Collaboration}
\affiliation{Corresponding Author: Ralf Ballhausen, \href{mailto:ralf.ballhausen@nasa.gov}{ralf.ballhausen@nasa.gov} \& Timothy R.\ Kallman, \href{mailto:timothy.r.kallman@nasa.gov}{timothy.r.kallman@nasa.gov}}

\author[0000-0003-4721-034X]{Marc Audard}
\affiliation{Department of Astronomy, University of Geneva, Versoix CH-1290, Switzerland} 

\author{Hisamitsu Awaki}
\affiliation{Department of Physics, Ehime University, Ehime 790-8577, Japan} 

\author[0000-0002-1118-8470]{Ralf Ballhausen}
\affiliation{Department of Astronomy, University of Maryland, College Park, MD 20742, USA} 
\affiliation{NASA / Goddard Space Flight Center, Greenbelt, MD 20771, USA}
\affiliation{Center for Research and Exploration in Space Science and Technology, NASA / GSFC (CRESST II), Greenbelt, MD 20771, USA}

\author[0000-0003-0890-4920]{Aya Bamba}
\affiliation{Department of Physics, University of Tokyo, Tokyo 113-0033, Japan} 

\author[0000-0001-9735-4873]{Ehud Behar}
\affiliation{Department of Physics, Technion, Technion City, Haifa 3200003, Israel} 

\author[0000-0003-2704-599X]{Rozenn Boissay-Malaquin}
\affiliation{Center for Space Sciences and Technology, University of Maryland, Baltimore County (UMBC), Baltimore, MD, 21250 USA}
\affiliation{NASA / Goddard Space Flight Center, Greenbelt, MD 20771, USA}
\affiliation{Center for Research and Exploration in Space Science and Technology, NASA / GSFC (CRESST II), Greenbelt, MD 20771, USA}

\author[0000-0003-2663-1954]{Laura Brenneman}
\affiliation{Center for Astrophysics | Harvard-Smithsonian, MA 02138, USA} 

\author[0000-0001-6338-9445]{Gregory V.\ Brown}
\affiliation{Lawrence Livermore National Laboratory, CA 94550, USA} 

\author[0000-0002-5466-3817]{Lia Corrales}
\affiliation{Department of Astronomy, University of Michigan, MI 48109, USA} 

% ...

\begin{abstract}
The X-ray binary system Cygnus X-3 (4U 2030+40, V1521 Cyg) is luminous but enigmatic owing to the high intervening absorption. High-resolution X-ray spectroscopy uniquely probes the dynamics of the photoionized gas in the system. In this paper we report on an observation of \cyg with the \xrism/Resolve spectrometer which provides unprecedented spectral resolution and sensitivity in the 2--10\,keV band. We detect multiple kinematic and ionization components in absorption and emission, whose superposition leads to complex line profiles, including strong P-Cygni profiles on resonance lines. The prominent \ion{Fe}{25} He$\alpha$ and \ion{Fe}{26} Ly$\alpha$ emission complexes are clearly resolved into their characteristic fine structure transitions. Self-consistent photoionization modeling allows us to disentangle the absorption and emission components and measure the Doppler velocity of these components as a function of binary orbital phase. We find a significantly higher velocity amplitude for the emission lines than for the absorption lines. The absorption lines generally appear blueshifted by ${\sim}-500\text{--}600\,\kms$. We show that the wind decomposes naturally into a relatively smooth and large scale component, perhaps originating with the background wind itself, plus a turbulent more dense  structure located close to the compact object in its orbit.
\end{abstract}

\section{Introduction}
\label{sec:intro}
\cyg is a high mass X-ray binary (HMXBs) with a very short orbital period \citep[$P = 4.8\,\mathrm{h}$;][]{Parsignault1972}. It is unusual in  its radio brightness, ranging from ${\sim}100\,\mathrm{mJy}$  up to 20\,Jy during outbursts \citep{Waltman1995}. Its distance is $9.7\pm0.5\,\mathrm{kpc}$ \citep{Reid2023}. The companion is a WN4-6 type Wolf-Rayet (WR) star \citep{vanKerkwijk1992, vanKerkwijk1996, Koljonen2017}. The strong wind from the companion star \citep{Paerels2000,Szostek2008a,Zdziarski2010,Zdziarski2012a} absorbs and scatters X-rays from the compact object, and produces a quasi-sinusoidal orbital light curve,  with its peak at phase 0.55--0.60 \citep{Willingale1985,Zdziarski2012a}. The wind material prevents unobscured observation of the compact X-ray source.

The X-ray line emission from \cyg is among the strongest of any known X-ray source \citep{Serlemitos1975}; with equivalent widths ranging from 0.2--0.4\,keV \citep{Hjalmarsdotter2008, Koljonen2010, Koljonen2018}. These lines can provide constraints on the conditions in the wind and other structures in the binary \citep{Paerels2000}. Study of these lines is aided by the brightness of \cyg, making it relatively well suited to study with current X-ray spectrometers. The spectrum is highly cut off by interstellar attenuation, with an average equivalent hydrogen column density of $\geq 10^{22}\,\mathrm{cm}^{-2}$ \citep{Kalberla2005}. Thus X-rays provide one of the only ways to study this system.  The X-ray line spectrum indicates the importance of photoionization as an excitation mechanism in the line emitting gas \citep[e.g.,][]{Liedahl1996}.  Observations using the Chandra High Energy Transmission Grating (HETG) have provided insights including the presence of radiative recombination continua (RRCs) which constrain the the gas temperature \citep{Paerels2000}, orbital modulation of the emission line centroids  \citep{Stark2003}, and P-Cygni profiles which constrain the masses of the two objects \citep{Vilhu2009,Kallman2019}. 
 
In this paper we report on the analysis of the  spectrum from \cyg as observed by the \xrism   X-ray observatory \citep{Tashiro2020}.  We address the following questions: What does the line spectrum look like when viewed with the high spectral resolution of \rsl?  What is the mechanism responsible for the  line emission, and what can we learn about the ionization state and other properties of the gas? What inferences can be made about the dynamics of the gas in the system?

\section{Data Acquisition and Reduction}
\label{sec:data}

\cyg was observed by \xrism as a Performance Verification (PV) target (starting 2024-03-24T06:53:50 UTC and ending at 2024-03-25T15:37:46 UTC) when the source was in a ``hypersoft state'', right before the onset of a major radio flare. %
The observation  resulted in a net exposure of ${\sim}66.6\,\mathrm{ks}$. The average flux 2--10\,keV was $4.7\times10^{-9}\,\mathrm{erg}\,\mathrm{cm}^{-2}\,\mathrm{s}^{-1}$, corresponding to an absorbed  luminosity of $5.3\times 10^{37}\,\mathrm{erg}\,\mathrm{s}^{-1}$ for an assumed distance of 9.7\,kpc \citep{Reid2023}. 

We used cleaned event files reprocessed by the Science Data Center (SDC) on 2024-07-24 and applied additional screening with respect to pulse rise time, event type, and status as recommended in the Quick-Start-Guide version 2.0. %
We further excluded events from pixel 27 due to known calibration uncertainties. The flat and point-source weighted composite resolutions of the whole array are both ${\sim}4.5\,\mathrm{eV}$ FWHM at 6\,keV.  The systematic uncertainty on the absolute energy scale from 5.4--9.0\,keV is $\pm0.3\,\mathrm{eV}$. We only use High Primary (Hp) event grades which give an average count rate of ${\sim}31\,\mathrm{cts}\,\mathrm{s}^{-1}$. We created standard exposure map, effective area, and response files with the \xrism SDC tools \texttt{xaarfgen} and \texttt{rslmkrmf} (Build 8 and CalDB 8) with default configuration. We use Large (L-type) response files for all spectral fitting as they describe the line-spread-function appropriately. Extra-large (XL-type) responses which include continuum redistribution effects lead to significant computational overhead and are not needed for narrow-band spectroscopy. 

The intensity variability associated with the orbital period is clearly visible in the \rsl lightcurve and the \xrism observation spans approximately seven orbital cycles of \cyg. We adopt the quadratic ephemeris of \cite{Antokin2019} with a slightly adjusted orbital period of 17252\,s \citep[${\sim}4.8\,\mathrm{h}$,][]{Kallman2019} which reproduces the widely accepted convention of phase zero to match the minimum of the orbital light curve (see also bottom panel of Fig.~\ref{phasefig2}).  We extract eight orbital phase resolved spectra of equal-sized width of 0.125 in phase. A multiplicity of the orbital period of \xrism and \cyg leads to periodic gaps in the light curve due to visibility constraints and renders phase bin 3 (0.250--0.375) severely underexposed and unsuitable for spectral analysis.

We used the data analysis packages \xspec \citep{Arnaud1996} and \isis \citep{Houck2000a} for spectral fitting. We re-binned the spectrum on a 2\,eV bin size grid and use $\chi^2$-statistics. Parameter uncertainties are statistical only and given at a $90\%$ confidence level unless otherwise noted. 

\section{Orbital phase-averaged spectroscopy}

\subsection{Empirical modeling and line identification}

The 2--10\,keV \rsl spectrum shows the expected multi-temperature blackbody with strong continuum absorption and a plethora of absorption and emission lines from various elements. In line with previous work \citep{Kallman2019, Koljonen2018} we model the 2--10\,keV continuum with a disk blackbody (\texttt{diskbb} in \xspec) with continuum absorption using the neutral absorption model  \texttt{tbabs} with \citet{Wilms2000} abundances and cross sections of \citet{Verner1995}. 

\begin{figure*}
    \centering
    \resizebox{\hsize}{!}{\includegraphics{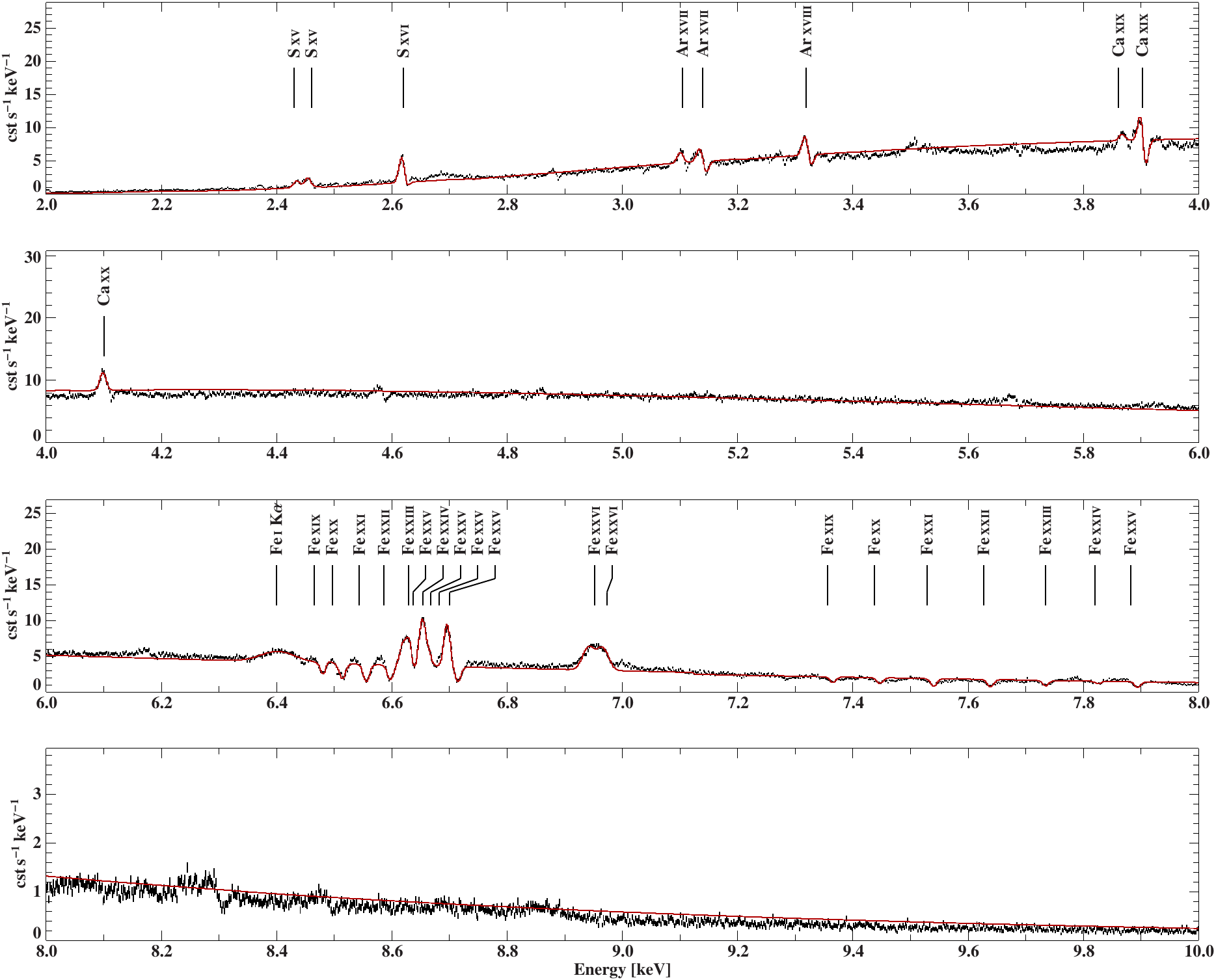}}
    \caption{2--10\,keV \rsl spectrum of \cyg, modeled with an absorbed disk blackbody and several Gaussian emission and absorption lines. Black labels are lines that are included in the model and listed in Table \ref{tab:gauss-table}.}
    \label{fig:rsl-broad-spec}
\end{figure*}

The broadband spectrum in Fig.~\ref{fig:rsl-broad-spec} clearly shows strong K-shell line emission and absorption from iron, near 6.4--8\,keV, as well as S, Ar, and Ca at lower energies.  The iron K$\alpha$ line complex consists of many narrow emission components separated by deep notches indicating absorption.

As a first step,  we fit the lines in the \rsl spectrum using Gaussians with centroids close to reference energies of probable transitions of each line. With the exception of the near-neutral Fe\,K$\alpha$ and the \ion{Fe}{26}\,Ly$\alpha$ lines, which required broader features, we kept the width of all lines fixed at 5\,eV. The results are given in Table~\ref{tab:gauss-table}. The line identification is tentative and often not unique as many of the observed features may be blends of unresolved fine structure transitions { or overlapping features of different ionization states. Our line labels therefore rather indicate plausible transitions}. We allow the normalization of each feature, which is proportional to the flux, and the apparent velocity shift to vary to obtain a best fit. Negative line normalizations indicate absorption and negative velocity shifts indicate blueshifted lines.  The fit model shown in Fig.~\ref{fig:rsl-broad-spec} is purely empirical and indeed not very satisfactory. This is mainly due to a practical limitation of the total number of Gaussian components that can be included in such a fit as well as the simplification of the profiles that often result from line blends. The purpose of this figure is rather to demonstrate these complications when inferring line velocity shifts and emphasize the importance of more detailed radiative transfer modeling.

\begin{deluxetable}{crrrrrrrrrr}
\label{tab:gauss-table}
\tablecaption{Emprical model lines \label{obstable}}
\tablewidth{0pt}
\tablehead{
\colhead{Energy}& Ion &\colhead{Transition}& \colhead{Reference}& \colhead{Norm} & Width &\colhead{$v_\mathrm{shift}$} \\
\colhead{keV} &  & &  & \colhead{$10^{-3}\,\mathrm{phs}\,\mathrm{s}^{-1}\,\mathrm{cm}^{-2}$} &  $\mathrm{eV}$& $\mathrm{km}\,\mathrm{s}^{-1}$
}
\startdata
2.430348 &    S\,\textsc{xv}    &  $1s^2\:^1S_0$--$1s^12p^1\:^3S_1$              & \xstar & $1.63\pm0.15$ & 5 & $-528^{+100}_{-97}$ \\  
2.460658 &    S\,\textsc{xv}    &  $1s^2\:^1S_0$--$1s^12p^1\:^1P_1$             & \xstar &  $2.25\pm0.16$ & 5 & $887^{+76}_{-72}$ \\
2.619679 &    S\,\textsc{xvi}   &  $1s^1\:^1S_{1/2}$--$1s^02p^1\:^2P_{1/2}$     & \xstar & $3.17\pm0.14$ & 5 & $436^{+28}_{-32}$  \\
3.104139 &    Ar\,\textsc{xvii} &  $1s^2\:^1S_0$--$1s^12p^1\:^3S_1$             & \xstar & $0.78\pm0.09$ & 5 & $385^{+57}_{-58}$ \\
3.139549 &    Ar\,\textsc{xvii} &  $1s^2\:^1S_0$--$1s^12p^1\:^1P_1$             & \xstar & $1.07^{+0.30}_{-0.12}$ & 5 & $456^{+107}_{-153}$ \\
3.139549 &    Ar\,\textsc{xvii} &  $1s^2\:^1S_0$--$1s^12p^1\:^1P_1$             & \xstar & $-0.82^{+013}_{-0.29}$ & 5 & $-470^{+133}_{-83}$ \\  
3.318176 &    Ar\,\textsc{xviii}&  $1s^1\:^1S_{1/2}$--$1s^02p^1\:^1P_{1/2}$     & \xstar & $1.37^{+0.37}_{-0.17}$ & 5 & $51^{+73}_{-94}$ \\ 
3.318176 &    Ar\,\textsc{xviii}&  $1s^1\:^1S_{1/2}$--$1s^02p^1\:^1P_{1/2}$     & \xstar & $-0.84^{+0.23}_{-0.66}$ & 5 & $-683^{+129}_{-112}$ \\
3.861134 &    Ca\,\textsc{xix}  &  $1s^2\:^1S_0$--$1s^12p^1\:^3S_1$             & \xstar & $-0.48^{+0.08}_{-0.09}$ & 5 & $-480^{+76}_{-82}$ \\ 
3.902255 &    Ca\,\textsc{xix}  &  $1s^2\:^1S_0$--$1s^12p^1\:^1P_1$             & \xstar & $2.03^{+1.91}_{-0.67}$ & 5 & $\leq188$ \\
3.902255 &    Ca\,\textsc{xix}  &  $1s^2\:^1S_0$--$1s^12p^1\:^1P_1$             & \xstar & $-2.04^{+0.68}_{-1.90}$ & 5 & $-268^{+97}_{-94}$ \\
4.100145 &    Ca\,\textsc{xx}   &  $1s^1\:^1S_{1/2}$--$1s^02p^1\:^2P_{1/2}$     & \xstar & $0.86\pm0.06$ & 5 & $110^{+39}_{-34}$ \\
6.400000 &    Fe\,\textsc{i}    &  K$\alpha$                                    & \xstar & $1.55\pm0.07$ & 30 & $-90\pm86$ \\
6.465592 &    Fe\,\textsc{xix}  & $2p^4\:^3P_2$--$1s^12s^22p^5\:^3P_2$          & \xstar & $-0.28^{+0.03}_{-0.04}$ & 5 & $-622^{+44}_{-35}$ \\
6.496421 &    Fe\,\textsc{xx}   & $2p^3\:^4S_{3/2}$--$1s^12s^22p^4\:^4P_{5/2}$  & \xstar & $-0.40\pm0.02$ & 5 & $-823^{+20}_{-24}$\\
6.496421 &    Fe\,\textsc{xx}   & $2p^3\:^4S_{3/2}$--$1s^12s^22p^4\:^4P_{5/2}$  & \xstar & $0.07^{+0.05}_{-0.04}$ & 5 & $207^{+168}_{-112}$ \\
6.543737 &    Fe\,\textsc{xxi}  & $2p^2\:^3P_0$--$1s^12s^22p^3\:^3D_1$          & \xstar & $-0.47\pm0.02$ & 5 & $-553^{+16}_{-15}$\\  
6.586495 &    Fe\,\textsc{xxii} & $2p^1\:^2P_{1/2}$--$1s^12s^22p^2\:^2D_{3/2}$  & \xstar & $-0.40\pm0.02$ & 5 & $-494^{+18}_{-17}$ \\    
6.628752 &    Fe\,\textsc{xxiii}& $2s^2\:^1S_0$--$1s^12s^22p^1\:^1P_1$          & \xstar & $-15^{+6}_{-46}$ & 5 & $-133^{+39}_{-31}$ \\ 
6.636579 &    Fe\,\textsc{xxv}  & $1s^2\:^1S_0$--$1s^12p^1\:^3S_1$~($z$)              & \xstar & $0.57\pm0.05$ & 5 & $784^{+28}_{-26}$ \\
6.661876 &    Fe\,\textsc{xxiv} & $2s^1\:^2S_{1/2}$--$1s^12s^12p^1\:^2P_{3/2}$  & \xstar & $0.41\pm0.04$ & 5 & $-30^{+59}_{-61}$ \\ 
6.667548 &    Fe\,\textsc{xxv}  & $1s^2\:^1S_0$--$1s^12p^1\:^3P_1$~($y$)             & \xstar & $16^{+13}_{-16}$ & 5 & $1617^{+38}_{-29}$ \\
6.682298 &    Fe\,\textsc{xxv}  & $1s^2\:^1S_0$--$1s^12p^1\:^3P_2$~($x$)              & \xstar & $1.18\pm0.05$ & 5 & $1348^{+13}_{-14}$ \\ 
6.700401 &    Fe\,\textsc{xxv}  & $1s^2\:^1S_0$--$1s^12p^1\:^1P_1$~($w$)             & \xstar & $1.15\pm0.04$ &  5 & $216^{+12}_{-10}$ \\
6.700401 &    Fe\,\textsc{xxv}  & $1s^2\:^1S_0$--$1s^12p^1\:^1P_1$~($w$)              & \xstar & $-0.40\pm0.02$ & 5 & $-630\pm18$ \\
6.951961 &    Fe\,\textsc{xxvi} & $1s^1\:^1S_{1/2}$--$1s^02p^1\:^2P_{1/2}$      & \xstar & $1.25^{+0.06}_{-0.09}$ & 10 & $465^{+50}_{-41}$ \\    
6.973174 &    Fe\,\textsc{xxvi} & $1s^1\:^1S_{1/2}$--$1s^02p^1\:^2P_{3/2}$      & \xstar & $1.16^{+0.07}_{-0.06}$ & 10 & $317^{+47}_{-38}$ \\  
7.356047 &    Fe\,\textsc{xix}  & $2s2.2p3\:^4S_{3/2}$--$1s2p^4(4P)3p\:^3P_{2}$    & \spex & $-0.16\pm0.02$ & 5 & $-377\pm45$ \\
7.437518 &    Fe\,\textsc{xx}   & $2s2.2p3\:^4S_{3/2}$--$1s2p^3(5S)3p\:^4P_{1/2}$  & \spex & $-0.18\pm0.02$ & 5 & $-366^{+47}_{-35}$ \\
7.528938 &    Fe\,\textsc{xxi}  & $2s^22p^2\:^3P_{0}$--$1s2p^2(4P)3p\:^3D_{1}$     & \spex & $-0.23\pm0.02$ & 5 & $-453^{+23}_{-29}$\\
7.627064 &    Fe\,\textsc{xxii} & $2s^22p\:^2P_{1/2}$--$1s2s^22p(3P)3p\:^4P_{1/2}$ & \spex & $-0.23\pm0.02$ & 5 & $-461^{+27}_{-26}$ \\
7.734248 &    Fe\,\textsc{xxiii}& $2s^2\:^1S_0$--$1s^12s^23p^1\:^1P_{1}$        & \spex & $-0.17\pm0.15$ & 5 & $-11\pm36$ \\
7.820462 &    Fe\,\textsc{xxiv} & $1s^22s^1\:^2S_{1/2}$--$1s^12s^1(1S)3p^1\:^2P_{3/2}$ & \spex & $-0.08\pm0.02$ & 5 & $-180^{+72}_{-66}$ \\
7.881520 &    Fe\,\textsc{xxv}  & $1s^2\:^1S_0$--$1s^13p^1\:^1P_1$              & \xstar & $-0.19\pm0.02$ & 5 & $-473^{+25}_{-30}$\\       
\enddata
\end{deluxetable}

Visual inspection of the spectrum as well as preliminary fitting of line features with Gaussians reveals both absorption and emission features over a wide range of ionization states of iron. Absorption and emission components accompany many strong lines, with the absorption generally blueshifted relative to the laboratory energy and the emission component redshifted. In this sense the line profiles resemble P-Cygni profiles \citep{beal29}, although the emission components of many lines are stronger than the absorption, indicating the importance of  processes such as electron impact excitation, recombination  or K shell fluorescence in addition to resonance scattering \citep{willis1986}. Among the most prominent features is the He$\alpha$ complex of \ion{Fe}{25} near 6.6--6.7\,keV, with a strong blueshifted absorption apparent in the resonance ($w$) line. The \rsl spectrum further shows a broad \ion{Fe}{26} Ly$\alpha$ emission feature near 7\,keV, emission from low-ionized iron near 6.4\,keV, and a series of K$\alpha$ and K$\beta$ absorption features from Li- through C-like iron.

It is clear from this analysis that no single red- or blueshift can fit to all the line centroids.  Most of the absorption components have Doppler blueshifts of ${\sim}500\,\kms$, while the redshifts of the emission components range from ${\sim}0\text{--}400\,\kms$ and no single value or narrow range of values is consistent with all of them.  This is difficult to reconcile with any simple picture of the motion of the gas in \cyg.

A solution to the puzzle of the line shifts comes from the fact that the absorption  and the emission overlap.  That is, the blueshifted absorption from a given line partially obscures the emission from adjacent emission lines. Even lines that do not form classic P-Cygni profiles due to their very low oscillator strength (for example He-like \ion{Fe}{25} $z$), can be affected by transitions of neighboring ionization states (here for example \ion{Fe}{24} $q$, $r$ and $t$), which makes a careful modeling of the whole ensemble of lines necessary to infer the dynamics of the absorbing and emitting gas. The exact contribution of these individual transitions to observed absorption and emission features requires appropriate radiative transfer modeling which we address in the following section. A decomposition of the photoionization model into emission and absorption components is shown in the lower two panels of Fig.~\ref{fig:warmabs-model}.

\subsection{Photoionization Modeling}
\label{modeling}

We have demonstrated that the interaction between emission and absorption  lines can account for the observed spectrum using detailed photoionization models which account for all the likely microphysical processes occurring in this gas, and utilize relatively recent state-of-the-art atomic energy levels, rate coefficients and cross sections.  We calculated ion populations for an optically thin medium with a density of $10^8\,\mathrm{cm}^{-3}$  and  solar metallicity \citep{ande89}  with an incident spectral energy distribution  of a multi-temperature blackbody (\texttt{diskbb} in \texttt{xspec)} with inner temperature of $kT =1.3\,\mathrm{keV}$, appropriate for this state of \cyg, using the plasma code \xstar v2.59\footnote{\url{https://heasarc.gsfc.nasa.gov/docs/software/xstar/xstar.html}}. The resulting charge state distribution (CSD) is overall similar to the one presented in \citet{Kallman2019}, although we note that they assumed a slightly hotter blackbody.  We have demonstrated that our results are unaffected by other abundance choices, including large hydrogen depletion as possibly expected in Wolf-Rayet stars.  We then used the ion populations with the \xstar-derived  \xspec/\isis analytic models \warmabs and \photemis to fit the narrow Fe\,K band spectrum (6--9\,keV) of \cyg. We kept the continuum absorption fixed as it cannot be constrained in the narrow band spectrum.  We find that a combination of three photoionized emission and three absorption components fit most of the Fe\,K$\alpha$ complex satisfactorily. We note that several Fe K$\beta$ lines are not well reproduced by the \xstar model alone, owing to incomplete treatment of turbulent broadening of these lines in \warmabs/\photemis, so we include an empirical Fe\,K$\beta$ model with reference energies as listed in Table~\ref{tab:gauss-table}. The motivation for this is that the Fe\,K$\beta$ lines appear mostly in absorption and therefore provide a useful constraint on the velocity of the absorption component.  Besides the blackbody normalization and temperature, the free parameters of our model are the local ionization parameter \logxi \citep{tart69}, the equivalent column density $N_\mathrm{H}$ for absorption models or normalization of emission models, respectively, as well as the turbulent broadening velocity $v_\mathrm{turb}$, and the bulk velocity shift $v_\mathrm{shift}$. In schematic \texttt{xspec} notation, our complete fit model can be written as $\texttt{tbabs}\times\left(\texttt{diskbb}+\texttt{vshift(\texttt{photemis}+\texttt{photemis}+\texttt{photemis})}\right)\times\texttt{vshift}(\texttt{warmabs}\times\texttt{warmabs}\times\texttt{warmabs}\times\texttt{Fe\,K}\beta)$.

The photoionized emission is modeled by one low ionization component ($\logxi\simeq1.6$) that accounts for most of the blended emission around 6.4\,keV, a moderately ionized component ($\logxi\simeq2.9$) with the charge state distribution peaking around \ion{Fe}{25} that reproduces the observed strong He$\alpha$ triplet lines, and a more highly ionized component ($\logxi\geq4.9$) with a charge state distribution consisting almost entirely of \ion{Fe}{26}, which gives rise to broad Ly$\alpha$ lines.  We tie the turbulent velocity broadening of the low and moderately high ionization emission components  whereas the highly ionized emission component is allowed to have a higher turbulent velocity.  Thus the \ion{Fe}{26} Ly$\alpha$ lines appear about twice as broad as the medium- and low-ionization lines. All emission components are redshifted by the same bulk velocity of ${\sim}130\,\kms$.

The emission components include the effects of radiative excitation by continuum photons, also referred to as `resonance scattering' \citep{kink02}.  This has the effect of boosting the strengths of essentially all resonance lines, most notably the $w$ line from He-like \ion{Fe}{25} \citep{gabe69}. The importance of this effect depends sensitively on the details of the radiative transfer at the line energy, and our treatment neglects attenuation of this radiation at the line energy, other than the continuum attenuation.  We have also experimented with assuming that all the radiation at the line energy is attenuated, i.e. we omit radiative excitation entirely.  This produces a much worse fit, owing to the fact that the remaining line excitation mechanisms, notably recombination, favor excitation of the intercombination and forbidden lines $x$, $y$, and $z$ of \ion{Fe}{25}, and hence under-produce the resonance line $w$. Similar results were hinted at in the  study of the third order \chandra/HETG spectra by \citet{sura24}. 

Our emission models intrinsically produce He-like line ratios appropriate to a low gas density, $10^8\,\mathrm{cm}^{-3}$.  The observed strengths of all the He-like lines are affected by overlapping line absorption, which complicates exploration of constraints on density or other processes affecting the relative strengths of the triplet series lines. Detailed modeling which examines the sensitivity of these effects to plasma conditions or dynamics are beyond the scope of this paper, and will be examined in subsequent publications.  

The absorption components reflect a similar separation of regions of different ionization: The low ionization component ($\logxi\simeq2.3$) reproduces a series of K$\alpha$ absorption lines of Li- through C-like Fe and a moderate ionization component  ($\logxi\simeq3.1$) gives rise to a strong \ion{Fe}{25} He$\alpha$~w line in absorption that forms a P-Cygni profile with the emission line. A highly ionized, $\logxi\geq4.5$ component is required to reproduce the full \ion{Fe}{26} Ly$\alpha$ profile through absorption. The absorption components have comparable column densities, i.e.  a flat absorption measure distribution (AMD) \citep{holc07}.  Again we tie the turbulent velocity broadening of the two lower ionized components and note that the absorption components generally appear to be roughly only half as velocity-broadened as the emission components and blueshifted with respect to their reference energies.  We do not favor a different arrangement of the absorbers and emitters in which the absorber does not absorb the emission.  Partial absorption of the emission lines is needed in order to account for the centroid locations for several of the emission lines.

Remaining residuals may be caused by the simplified line profile modeling that assumes turbulent broadening but neglects the bulk velocity dispersion in the WR wind. Other potential model short-comings may include insufficient broadening of the Fe K$\beta$ model lines. Line centroids, however, are reproduced satisfactorily, allowing for detailed velocity measurements, as well as the depth of the majority of lines.   We have also experimented with allowing for velocity shifts of individual components to vary independently. This does not  improve the fits nor does  it significantly affect the values of the shifts.

\begin{figure*}[h]
   \centering
    \resizebox{\hsize}{!}{\includegraphics{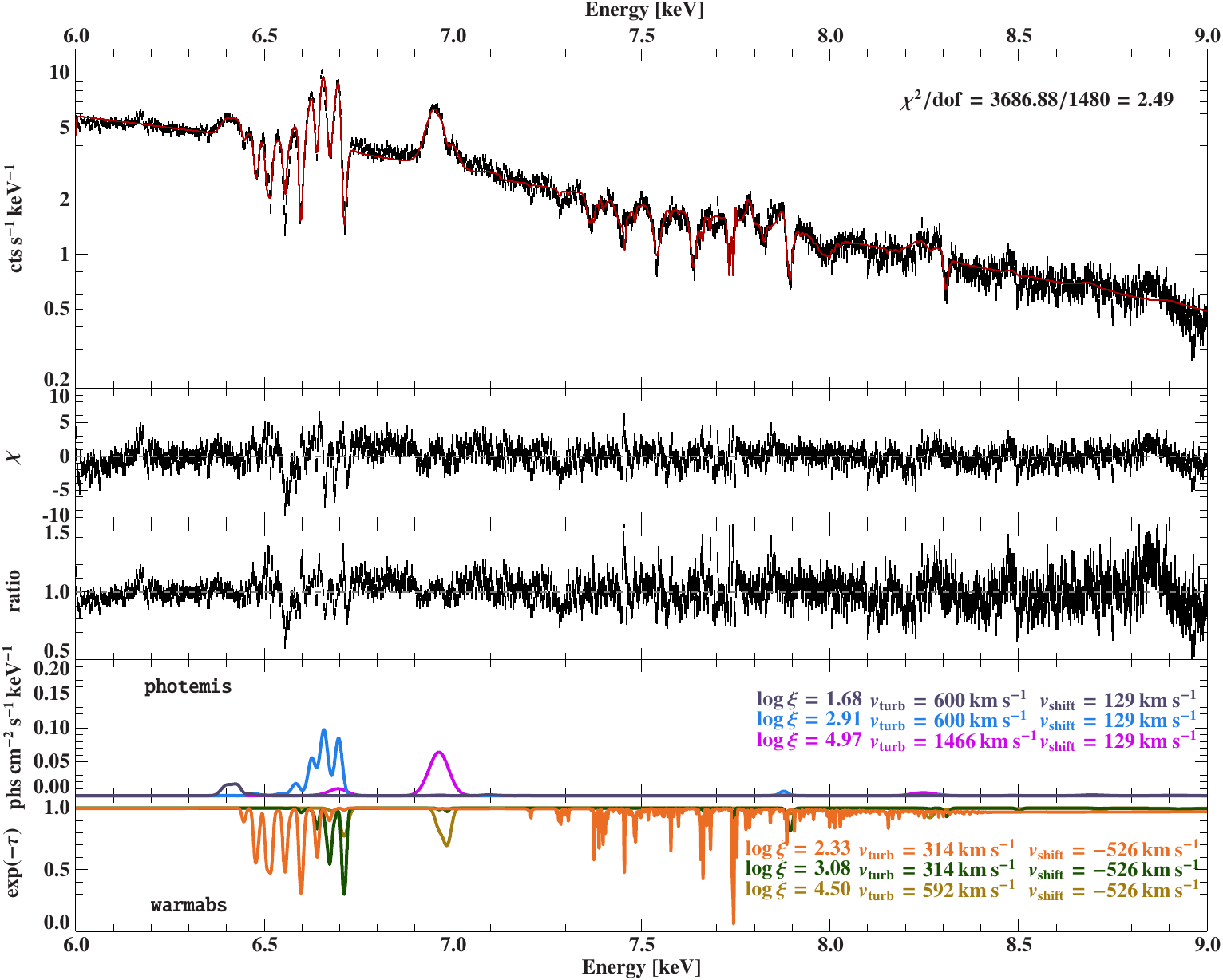}}
    \caption{The \rsl spectrum in the Fe K region showing the observed data and our best-fit model using multiple photoionized absorption and emission components as described in the text.  The top panel shows the observed count data (black) together with the best fit model.  The second and third panels show the contributions to $\chi^2$ vs. energy and the ratio of the best fit model to the observed counts.  The final two panels show the models for line emission and absorption, respectively.  Note the energy offset between the emission and absorption. }
    \label{fig:warmabs-model}
\end{figure*}

Our best fit is shown in Fig.~\ref{fig:warmabs-model} and parameters are given in Table~\ref{tab:orbavg-parameters}. The lower two panels of Fig.~\ref{fig:warmabs-model} show the respective contributions of the individual model components. 

\begin{deluxetable}{lr|rrrrrrr}
\label{tab:orbavg-parameters}
\tabletypesize{\scriptsize}
\tablecaption{Best-fit parameters of the photoionization model of the phase-averaged and phase-resolved spectra.}
\tablewidth{0pt}
\tablehead{
\colhead{Parameter}& \colhead{phase-average} &\colhead{Bin 1}& \colhead{Bin 2}& \colhead{Bin 4} & 
\colhead{Bin 5}&\colhead{Bin 6} &
\colhead{Bin 7} & \colhead{Bin 8} 
}
\startdata
        \multicolumn{9}{c}{Continuum}\\
        \hline
        $\log\left(N_\mathrm{H}\right)$\tablenotemark{c}                        & $22.72^{\dagger}$          & $22.72^{\dagger}$         & $22.72^{\dagger}$         & $22.72^{\dagger}$         & $22.72^{\dagger}$         & $22.72^{\dagger}$         & $22.72^{\dagger}$         & $22.72^{\dagger}$          \\
        $\mathrm{Norm}_\texttt{diskbb}$                                         & $301^{+10}_{-8}$           & $364^{+39}_{-34}$         & $408^{+29}_{-27}$         & $425^{+35}_{-28}$         & $448^{+38}_{-36}$         & $661^{+79}_{-67}$         & $265\pm^{+21}_{-19}$      & $295^{+39}_{-35}$          \\
        $T_\mathrm{in}$~[$\mathrm{keV}$]                                        & $1.30\pm0.01$  & $1.26\pm0.02$ & $1.25\pm0.02$           & $1.24\pm0.02$ & $1.23\pm0.02$           & $1.15\pm0.02$           & $1.31\pm0.02$               & $1.26^{+0.03}_{-0.01}$  \\
        \hline
        \multicolumn{9}{c}{photemis  component \#1}\\
        \hline
        $\log(\mathrm{norm})$                                                   & $4.82^{+0.03}_{-0.07}$ & $4.97^{+0.07}_{-0.08}$ & $4.97^{+0.06}_{-0.08}$ & $4.74^{+0.08}_{-0.10}$  & $4.81^{+0.09}_{-0.11}$ & $4.83^{+0.13}_{-0.21}$ & $4.82\pm{0.06}$ & $5.046^{+0.063}_{-0.096}$ \\
        $\log\xi$                                                               & $1.68\pm0.08$             & $1.58^{+0.03}_{-0.02}$  & $1.57\pm0.2$              & $1.69^{+0.02}_{-0.01}$     & $1.58\pm0.03$             & $1.59^{+0.10}_{-0.04}$     & $1.68\pm0.01$             & $1.58^{+0.03}_{-0.02}$    \\
        $v_\mathrm{turb}$\tablenotemark{a}~[$\mathrm{km}\,\mathrm{s}^{-1}$]     & $600^{+26}_{-21}$         & $661^{+68}_{-42}$         & $703^{+49}_{-57}$         & $503^{+59}_{-48}$          & $528^{+33}_{-32}$         & $604^{+89}_{-95}$          & $631\pm50$                &  $726^{+133}_{-63}$         \\
        \hline
        \multicolumn{9}{c}{photemis component \#2}\\
        \hline
        $\log(\mathrm{norm})$                                                   & $3.43\pm{0.03}$ & $3.40^{+0.07}_{-0.05}$ & $3.34^{+0.05}_{-0.03}$ & $3.37\pm0.07$            & $3.45^{+0.02}_{-0.04}$ & $3.51^{+0.09}_{-0.07}$ & $3.54^{+0.03}_{-0.04}$ & $3.68\pm{0.04}$ \\
        $\log\xi$                                                               & $2.91^{+0.02}_{-0.06}$    & $2.94^{+0.09}_{-0.02}$    & $2.94\pm0.02$             & $2.83^{+0.08}_{-0.07}$     & $2.89^{+0.05}_{-0.06}$    & $2.92^{+0.04}_{-0.09}$    & $2.91\pm0.02$             & $2.91^{+0.07}_{-0.08}$    \\
        $v_\mathrm{turb}$\tablenotemark{a}~[$\mathrm{km}\,\mathrm{s}^{-1}$]     & $600^{+26}_{-21}$         & $661^{+68}_{-42}$         & $703^{+49}_{-57}$         & $503^{+59}_{-48}$          & $528^{+33}_{-32}$         & $604^{+89}_{-95}$         & $631\pm50$                & $726^{+133}_{-63}$         \\  
        \hline
        \multicolumn{9}{c}{photemis component \#3}\\
        \hline
        $\log(\mathrm{norm})$                                                   & $2.21^{+0.02}_{-0.03}$ & $2.13^{+0.06}_{-0.03}$ & $2.09\pm{0.03}$ & $2.12\pm{0.06}$  & $2.18\pm{0.03}$ & $2.52^{+0.11}_{-0.08}$ & $2.42^{+0.04}_{-0.03}$ & $2.59^{+0.06}_{-0.09}$ \\
        $\log\xi$                                                               & $\ge4.92$                 & $\ge4.85$                 & $4.65^{+0.12}_{-0.05}$    & $4.63^{+0.22}_{-0.09}$     & $\ge 4.90$                & $\ge 4.89$                & $\ge 4.94$                & $\ge 4.55$                \\ 
        $v_\mathrm{turb}$~[$\mathrm{km}\,\mathrm{s}^{-1}$]                      & $1466\pm52$               & $1677^{+158}_{-206}$       & $\ge1919$                 & $\ge 1889$                 & $1378^{+132}_{-117}$      & $1180^{+92}_{-98}$        & $1211^{+49}_{-57}$        & $1070^{+75}_{-69}$        \\
        \hline
        \multicolumn{9}{c}{warmabs component \#1}\\
        \hline
        $\log\left(N_\mathrm{H}\right)$                                         & $22.73\pm0.02$            & $22.88\pm0.04$            & $22.82\pm0.02$            & $22.70\pm0.04$             & $22.66\pm0.03$             & $22.70^{+0.05}_{-0.07}$   & $22.698^{+0.04}_{-0.03}$    & $22.75^{+0.05}_{-0.06}$            \\
        $\log\xi$                                                               & $2.33\pm{0.01}$           & $2.19\pm0.04$             & $2.32^{+0.01}_{-0.02}$    & $2.37\pm0.02$              & $2.37\pm0.02$              & $2.33^{+0.02}_{-0.03}$    & $2.32^{+0.01}_{-0.03}$    & $2.20^{+0.06}_{-0.02}$   \\
        $v_\mathrm{turb}$\tablenotemark{b}~[$\mathrm{km}\,\mathrm{s}^{-1}$]     & $313^{+11}_{-21}$         & $401^{+35}_{-32}$         & $365^{+16}_{-20}$         & $322^{+23}_{-25}$          & $320^{+19}_{-26}$           & $280^{+50}_{-53}$         & $260^{+23}_{-34}$         & $223^{+49}_{-52}$        \\
        \hline
        \multicolumn{9}{c}{warmabs component \#2}\\
        \hline
        $\log\left(N_\mathrm{H}\right)$                                         & $22.05^{+0.05}_{-0.06}$   & $22.15^{+0.07}_{-0.08}$   & $22.29^{+0.03}_{-0.06}$   & $22.29\pm0.07         $   & $22.35^{+0.07}_{-0.09}$    & $22.37^{+0.07}_{-0.08}$   & $21.95^{+0.09}_{-0.10}$    & $21.64^{+0.15}_{-0.24}$ \\
        $\log\xi$                                                               & $3.08^{+0.08}_{-0.04} $   & $3.31^{+0.13}_{-0.06}$    & $3.13^{+0.10}_{-0.03}$    & $3.04^{+0.02}_{-0.03}$    & $3.02\pm0.03$              & $3.05^{+0.06}_{-0.04}$    & $3.04^{+0.04}_{-0.05}$    & $3.04^{+0.23}_{-0.21}$  \\
        $v_\mathrm{turb}$\tablenotemark{b}~[$\mathrm{km}\,\mathrm{s}^{-1}$]     & $313^{+11}_{-21}$         & $401^{+35}_{-32}$         & $365^{+16}_{-20}$         & $322^{+22}_{-28}$         & $320^{+19}_{-26}$          & $280^{+50}_{-53}$         & $260^{+23}_{-34}$         & $223^{+49}_{-52}$       \\
        \hline
        \multicolumn{9}{c}{warmabs component \#3}\\
        \hline
        $\log\left(N_\mathrm{H}\right)$ \tablenotemark{d}                       & $22.35^{+0.15}_{-0.08}$   &   ---                     & ---                       & $22.28^{+0.15}_{-0.21}$   & $22.66^{+0.06}_{-0.07}$    & $23.00^{+0.10}_{-0.15}$   & $22.68^{+0.07}_{-0.05}$    & $22.85^{+0.08}_{-0.10}$ \\
        $\log\xi$                                                               & $4.50^{+0.16}_{-0.10}$    &    ---                    & ---                       & $\ge 4.62$                & $\ge 4.90$                 & $4.66^{+0.20}_{-0.12}$    & $4.46^{+0.07}_{-0.04}$     & $4.33^{+0.13}_{-0.07}$ \\
        $v_\mathrm{turb}$~[$\mathrm{km}\,\mathrm{s}^{-1}$]                      & $592^{+52}_{-58}$         &    ---                    & ---                       & $300^{+105}_{-92}$        & $394^{+82}_{-78}$           & $\ge 727$                 & $734^{+66}_{-60}$          & $\ge 937$              \\
        \hline
        $v_\mathrm{photemis}$~[$\mathrm{km}\,\mathrm{s}^{-1}$]                  & $128^{+17}_{-35}$         & $-22\pm40$                & $-156^{+43}_{-40}$        & $32^{+56}_{-67}$          & $68^{+630}_{-67}$           &  $135^{+74}_{-92}$       & $239^{+26}_{-24}$          & $152^{+60}_{-59}$      \\
        $v_\mathrm{warmabs}$~[$\mathrm{km}\,\mathrm{s}^{-1}$]                   & $-526\pm6$                & $-600\pm30$               & $-576^{+10}_{-28}$        & $-494^{+14}_{-7}$        & $-496^{+17}_{-11}$         &  $-476^{+22}_{-18}$       & $-525^{+12}_{-6}$          & $-566^{+24}_{-11}$     \\
        \hline
        $\chi^2/\mathrm{dof}$                                                   &  2.49                     & 1.30                      & 1.39                      & 1.27                     & 1.39                       & 1.24                       & 1.48                       & 1.24                   \\
      \hline\hline
      \enddata
      \tablenotetext{a}{These parameters are tied together during spectral fitting.}
      \tablenotetext{b}{These parameters are tied together during spectral fitting.}
      \tablenotetext{c}{Units of $N_\mathrm{H}$ in this table are $\log_{10}$ of the column density in units of $\mathrm{cm}^{-2}$.}
      \tablenotetext{d}{The column density of the third, highest ionized absorption component (\warmabs \#3) that primarily models the \ion{Fe}{26} Ly$\alpha$ absorption lines hit the lower limit of $10^{20}\,\mathrm{cm}^{-2}$ in the spectra of Bin 1 and 2 and was therefore removed in the final fits.}
      \tablenotetext{\ensuremath{\dagger}}{This parameter was fixed during spectral fitting.}
\end{deluxetable}

\section{Orbital phase-resolved spectroscopy}

\begin{figure}
    \centering
    \resizebox{\hsize}{!}{\includegraphics{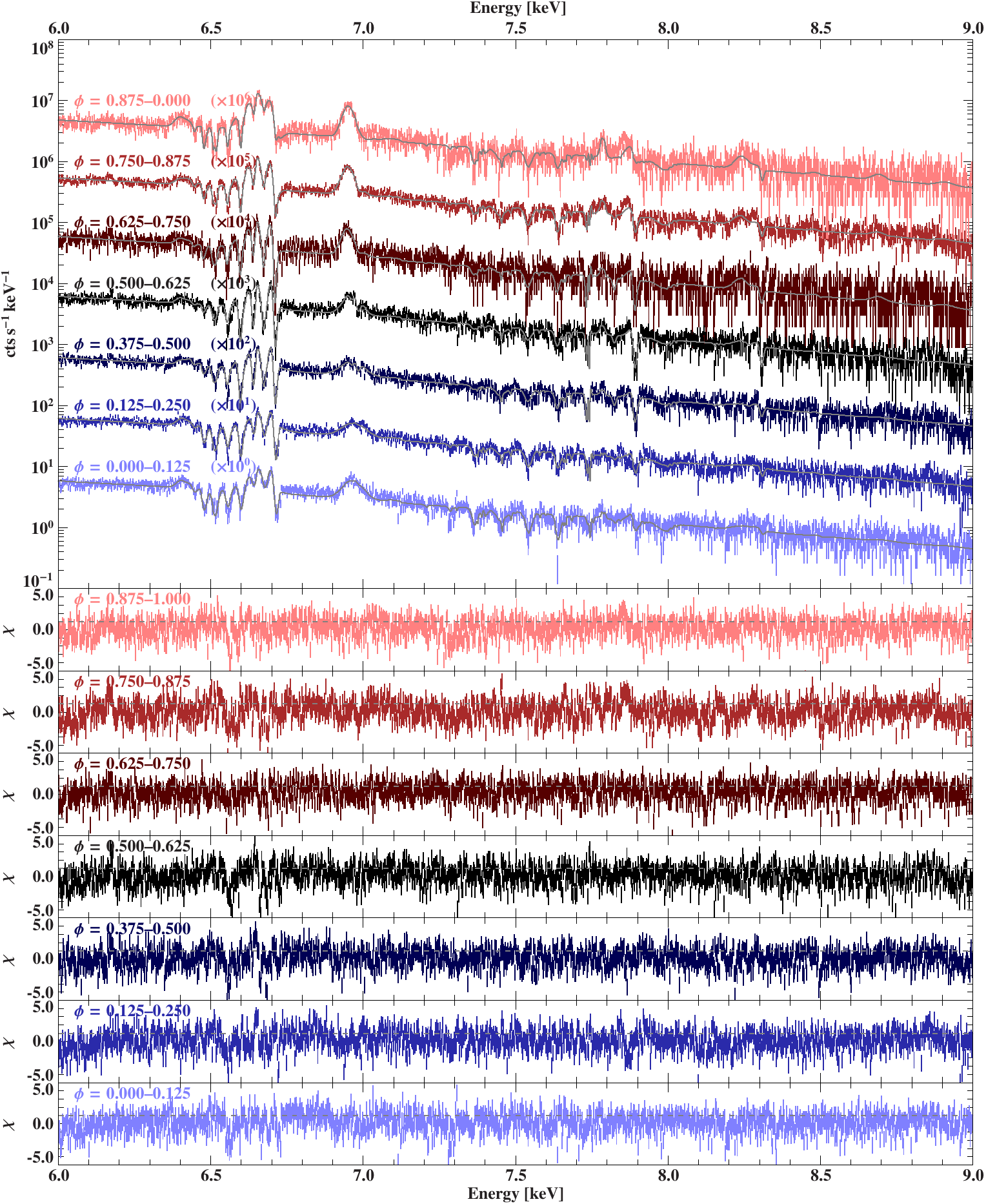}}
    \caption{Top: Orbital phase-resolved \rsl spectra of \cyg. The gray line shows the respective best-fit model which has the same components as the phase-averaged model described in Sect.~\ref{modeling}. Bottom: $\chi$-residuals} of the individual phase-resolved spectra.
    \label{fig:orb-ratios}
\end{figure}

\begin{figure*}[!]
    \centering
     \includegraphics[angle=0, width=1\textwidth]{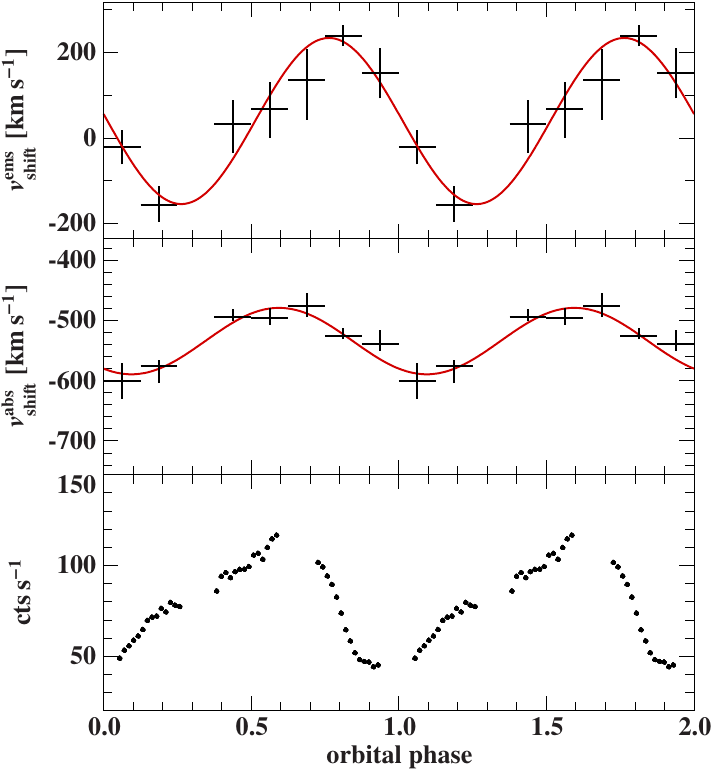}
   \caption{Top: Velocity shift of the photoionized emission components of the spectrum as a function of orbital phase. Middle:  Velocity shift of the photoionized absorption components of the spectrum as a function of orbital phase. Red curves show fits of the radial velocity profile (see text for details). Bottom: 2--10\,keV \rsl light curve folded on the orbital period of 17252\,s. Two orbital cycles are shown for clarity.}
   \label{phasefig2}
\end{figure*}

We have explored the variability of the spectrum with orbital phase.  We divided the data into 8 equally spaced phase bins as described in Sect~\ref{sec:data}. %
The procedure for fitting is the same as in section \ref{modeling} and we again only focus on the 6--9\,keV band.  Figure~\ref{fig:orb-ratios}  shows the spectrum in the various phase bins. The lower panels of Fig.~\ref{fig:orb-ratios} show the $\chi$ residuals of each phase-resolved spectrum with respect to the best-fit model.  The best-fit parameters for each of the orbital phase-resolved spectra are also listed in Table~\ref{tab:orbavg-parameters} and we show the variation of the velocity shift of the absorption and emission components as a function of orbital phase in Fig.~\ref{phasefig2}. We fitted for the systemic velocity $v_\mathrm{sys}$ and the projected velocity semi-amplitude $K$ for both components assuming strictly circular orbits (i.e., an eccentricity $e=0$). We find that the emission component has a systemic velocity of $v_\mathrm{sys}^\mathrm{ems}= 40\pm20\,
\kms$ and a semi-amplitude $K^\mathrm{ems}=194\pm29\,\kms$ while  the absorption component has a systemic velocity $v_\mathrm{sys}^\mathrm{abs}= -534\pm6\,\kms$ with a semi-amplitude of  $K^{abs}=55\pm7\,\kms$. We note that these velocities are similar to the velocity shifts for the orbital phase-averaged spectrum reported in Table~\ref{tab:orbavg-parameters}, although the different weighting of the individual phase bins leads to slightly different values. We
observe the maximum redshift of the emission and absorption components at phase $0.75\pm0.03$ and $0.60\pm0.03$, respectively. 

Examination of Table~\ref{tab:orbavg-parameters}  and Fig.~\ref{phasefig2} shows the following:

\begin{itemize}
    \item{The phase modulation of the 2--10\,keV folded light curve  is not sinusoidal. While the folded light curve has per definition a minimum at phase 0.0,  the maximum occurs near phases 0.6--0.7, rather than at phase 0.5. Previous observations have shown that the long-term phase-folded continuum is also not sinusoidal, and peaks at phase 0.55--0.60 \citep{Antokin2019, Zdziarski2012a}.}
    \item{ The emission components have a velocity variation of
    $K = 194 \pm 29\,\kms $ and an approximately sinusoidal phase dependence.  The radial velocity of the average flux is $v_\mathrm{sys}^\mathrm{ems}=40\pm 20\,\kms $. The extreme values occur near phases 0.25 and 0.75.}
    \item{The intensity of these emission components varies by a factor of $\simeq$2 around the orbit, with the maximum value occurring near phase 0.75, following a  trend similar to the broadband intensity profile.}  
    \item{The ionization parameters of the emission components do not vary significantly around the orbit. All phase-resolved spectra show strong lines of the same ion species, except the \ion{Fe}{26} Ly$\alpha$ absorption lines which are very weak in Bins 1 and 2 (i.e. phases 0--0.25).}
    \item{The absorber velocities vary with the orbit with an amplitude of  $K=55 \pm 7\,\kms$, an average value of $v_\mathrm{sys}^\mathrm{abs} = -534 \pm 6\,\kms$ and show extreme values at phases $\sim$0.1 and $\sim$0.6.}
    \item{The amplitude of the velocity modulation of the emission components is larger than its average, so in some phase bins it appears redshifted while in others it appears blueshifted. The absorption components are always blueshifted by at least $-400\,\kms$.}
\end{itemize}

Although the photoionization model reproduces the overall spectra well, we note distinct characteristics of the \ion{Fe}{26} Ly$\alpha$ emission in both the phase-averaged and phase-resolved spectra. The width of this line, both in emission and absorption, is about twice as large as for all other lines and the absorption line depths do not follow the same trend as ionization states \ion{Fe}{25} and below. For this reason, we have also modeled the line profile of the \ion{Fe}{26} Ly$\alpha$ separately, to provide a better comparison with other lines.

We fit \ion{Fe}{26}\,Ly$\alpha$ line in the spectra of  Bins 2, 4, and 5, where evidence of absorption features is confirmed, with a model consisting of four \xspec \texttt{zgauss} components and a single \texttt{powerlaw} continuum: two \texttt{zgauss} for the Ly$\alpha_1$ and Ly$\alpha_2$ emission and the other two are the Ly$\alpha_1$ and Ly$\alpha_2$ absorption (and thus the latter two have negative normalization).  We fix the Ly$\alpha_1$ and Ly$\alpha_2$ line energies to their rest-frame energies (6973\,eV and 6952\,eV, respectively), and the Ly$\alpha_1$/Ly$\alpha_2$ flux ratio to 2. The redshift and  width of the Ly$\alpha_1$ and Ly$\alpha_2$ lines are assumed to be identical within the emission or absorption component, but are independent between the emission and absorption.  We obtain $v_{\rm shift} = -392^{+111}_{102}$\,\kms (Bin 2), 
$-81_{-48}^{+37}$\,\kms (Bin 4), and 
$73_{-305}^{+244}$\,\kms (Bin 5) for the emission component, and 
$-397_{-130}^{+152}$\,\kms (Bin 2), 
$-433 \pm 56$\,\kms (Bin 4), and 
$-367_{-106}^{+79}$\,\kms (Bin 5) for the absorption component.
This indicates a larger amplitude of the orbital modulation in the emission component than in the absorption components.  This is qualitatively consistent with the results from the \xstar-based photoionization model, but the amplitude obtained here for the emission component is significantly larger. This suggests that the highest ionization component is spatially separate from the low- to medium ionization components of the emission, and thus is kinematically separate as well.  More details of the analysis focusing on the \ion{Fe}{26} Ly$\alpha$ lines will be presented in a separate paper.

\section{Discussion}
\label{sec:discussion}
\subsection{Velocities}

We measured the orbital velocity curves of the absorption and emission features in \cyg by fitting our multi-component photoionization model to the phase-resolved Fe\,K band spectra. In this model, velocity shifts are tied such that the parameter space is reduced to a single velocity shift for all absorption features and a single velocity shift for all emission features. We note that in this model the velocity shift of the absorption features is mostly constrained by the series of absorption lines from \ion{Fe}{20} through \ion{Fe}{25} while the emission component is constrained primarily by the \ion{Fe}{25}\,He$\alpha$ emission complex. The \ion{Fe}{26} Ly$\alpha$ lines generally appear much broader and absorption lines on top of the strong emission feature are present only in some phase intervals. This behavior suggests that these lines may probe a different region of the stellar wind with possibly different dynamics. We find that our photoionization model generally describes the profile of the \ion{Fe}{26} well if we allow for a larger turbulent broadening of these lines, a result that is supported by our empirical Gaussian model.

The overall small systemic velocity and very sinusoidal modulation of the velocity shift of the emission components suggest that their origin is in the vicinity of the compact object and the velocity modulation being closely connected to its orbit. Further confirmation for this comes from the orbital phase variation:  the extreme values of the radial velocity of the emission occur at phases 0.25 and 0.75.  This is just what would be expected for a circular orbit with inferior conjunction at phase 0.  Further confirmation of this hypothesis comes from the fact that the data require that the absorption absorbs the emission, which indicates that the absorber must be further from the compact object than the emitting gas.

We note parenthetically that the relation between the location of the emitting gas and its ionization parameter depends strongly on the density distribution close to the compact object.  It is likely that the density will be influenced by the stellar wind together with structures affected by the  gravity of the compact object: accretion wake, bow shock, or accretion disk  \citep[e.g.,][]{blon91,elme19}.  Simple assumptions about the gas density are therefore highly uncertain.

The interpretation of the velocity modulation of the absorption component is less obvious. The absorption components have a  systemic blueshift of ${\sim}-530\,\kms$ with a semi-amplitude ${\sim}55\,\kms$.  Such semi-amplitude velocities are much less than the corresponding values for the IR emission lines, which are attributed to the primary star \citep{Koljonen2017}. If so, it is unclear why the X-ray lines do not show a comparable speed.

Simple estimates suggest that the ionization parameter of the stellar wind illuminated by the X-rays from the compact object is $\log\xi{\sim}$3--4 (see also the discussion in section~\ref{sec:diss_abslines}), which is comparable to the values found for absorption components 1 and 2 in Table~\ref{tab:orbavg-parameters} and that such ionization parameters can be sustained over large fractions of the wind.  This suggests that the absorption is due to gas on a spatial scale large with respect to the binary orbit, with density comparable to the primary stellar wind, but with dynamics which are closer to being symmetric around the system center of mass rather than being tied to the orbital motion of the primary star. More detailed inferences about this gas require testing of specific dynamical models, which we will carry out in a subsequent paper.

The velocity profile of the absorption lines is more difficult to interpret. The line of sight is expected to cross large volumes of the wind and the initial speed as well as the velocity  dispersion and the density profile of the wind will contribute to the observed absorption line profile and its orbital dependence. The efficiency of the wind acceleration is further subject to the degree of ionization by the compact object. While our fitted semi-amplitude of the velocity of the absorption lines is ${\sim}55\,\kms$ we note that the change in turbulent velocity along the orbit is also of the order of several tens of \kms for \warmabs components 1 and 2, emphasizing the complexity of the wind dynamics and caveats to the interpretation of the observational results.

It is still instructive to discuss quantitative consequences of our results based on plausible assumptions. An estimate for the orbital separation is $3.4\,R_\odot = 2.4\times10^{11}\,\mathrm{cm}$ \citep{Vilhu2009}. For wind speeds of a few hundred \kms (before the wind reaches its terminal velocity), the time for perturbations of the wind velocity to propagate through the binary system is on the order $\sim 10^3$\,s.  This can be compared with the orbital period of $\sim10^4$\,s.  This suggests that the wind structure will reflect the motion of the primary star out to large distances.    However, significant apparent phase shifts of the orbital velocity curve are still possible, and this could account for the small phase shift of the observed absorption line radial velocity curve.

If we assume the semi-amplitudes of the absorption and emission components are representative of the motion of the primary star and the compact object, respectively, then they allow to constrain the mass ratio $q=M_\mathrm{WR}/M_\mathrm{CO}$,  which we find to be $3.8\pm0.6$ using standard Gaussian error propagation. Deriving absolute masses is  sensitive to the inclination of the system. A plausible range of nitrogen-rich Wolf-Rayet star masses is 10--83\,$M_\odot$\citep{Crowther2007}.

 More detailed inferences about the system masses and parameters are complicated owing to the dependence on inclination, assumptions about the primary, and assumptions about the locations of the various emission components in the binary system.   We believe that the \xrism spectrum provides a new opportunity to address these issues, and our model fits are the first step toward new constraints on the masses.   However, a thorough examination of these issues requires extensive  modeling and analysis of the 3D dynamics and radiative transfer of the \cyg binary system and we have elected to postpone this until a future paper.

\subsection{Emission line luminosities}

The model parameters allow us to infer the conditions in the emitting and absorbing gas in \cyg.  Fits of the emission components provide a normalization for the model, as shown in Table~\ref{tab:orbavg-parameters}. For the \photemis model\footnote{For details, see the \xstar manual \url{https://heasarc.gsfc.nasa.gov/xstar/xstar.html}.}, the normalization $\kappa$ is directly related to the emission measure 
\begin{equation}
    EM = 1.20\times10^{54} \times \kappa \times \left(\frac{D}{1\,\mathrm{kpc}}\right)^2\,\mathrm{cm}^{-3}~,
\end{equation}
where $D$ is the distance to the source.
Assuming a distance of 9.7\,kpc \citep{Reid2023}, \photemis component 2 corresponds to an emission measure  $EM=3.0\times10^{59}\,\mathrm{cm}^{-3}$.  This can be compared with the value derived from the emissivity of the \ion{Fe}{25} resonance line ($w$).  At $\xi=10^3\,\mathrm{erg}\,\mathrm{cm}\,\mathrm{s}^{-1}$ the emissivity of this line we obtain from \xstar is $j/n^2=7.91 \times 10^{-25}$\,erg\,cm$^3$\,s$^{-1}$.   From the Gaussian fits we infer a luminosity of $L_\mathrm{line}=1.39 \times 10^{35}\,\mathrm{erg}\,\mathrm{s}^{-1}$, so the emission measure by this method is $EM=1.76 \times 10^{59}\,\mathrm{cm}^{-3}$. We note that this is at best a rough estimate, since the line emissivity is sensitive to assumptions about density, the rate for radiative excitation and the charge state distribution. This emission measure corresponds to the total density of nuclei, not just iron, and it assumes solar abundances of iron.  WR winds are expected to be highly depleted in hydrogen.  This affects the ionization balance by resulting in a greater electron number density, relative to the nuclei. This can enhance the total line emissivity by factor $\leq$2.

The emission measures derived above can be compared with the values which can be provided by a stellar wind appropriate for a WR star. If the star has a mass loss rate $\dot{M}=10^{-5}\,M_\odot\,\mathrm{yr}^{-1}$ and wind speed $v$ of 1000\,\kms (we assume it to be constant for now), then the wind density as a function of radius is 
\begin{equation}
n(R)=\frac{\dot{M}/\mu m_\mathrm{H}}{4 \pi R^2 v}=3.04 \times 10^{11}\,\mathrm{cm}^{-3}\left(\frac{R}{10^{12}\,\mathrm{cm}} \right)^{-2} \left(\frac{v}{1000\,\mathrm{km}\,\mathrm{s}^{-1}}\right)^{-1}\left( \frac{\dot{M}}{10^{-5}\,M_\odot\,\mathrm{yr}^{-1}}\right)~,
\end{equation}
where $\mu$ is the average atomic weight per nucleon (here we assume $\mu$=1, i.e., pure hydrogen). The total emission measure in the wind is then obtained by integration from an innermost radius $R$ to infinity:
\begin{equation}
EM=\frac{1}{4 \pi R} \left(\frac{\dot{M}/\mu m_\mathrm{H}}{v}\right)^2=1.16 \times 10^{60}\,\mathrm{cm}^{-3} \left(\frac{R}{10^{12}\,\mathrm{cm}} \right)^{-1} \left(\frac{v}{1000\,\mathrm{km}\,\mathrm{s}^{-1}}\right)^{-2}\left( \frac{\dot{M}}{10^{-5}\,M_\odot\,\mathrm{yr}^{-1}}\right)^2~.
\end{equation}

This estimate is  greater than what is needed to produce the observed lines assuming the wind is optically thin, based on the simple estimates given above.   Thus it is plausible that the wind itself can supply the gas seen in the X-ray lines.  More detailed comparisons must include models or assumptions about the gas dynamics together with detailed multi-dimensional radiative transfer calculations

\subsection{Absorption lines}
\label{sec:diss_abslines}
The results of our spectal fitting presented in Sect.~\ref{modeling} show the presence of ionized plasma components with an ionization parameter of order $\log\xi\simeq2\text{--}3$. This is consistent with crude order of magnitude estimates of the ionizing conditions in a system with  column density $\simeq 10^{22}\,\mathrm{cm}^{-2}$, a density $n\sim 10^{11}\,\mathrm{cm}^{-3}$, size of the absorbing medium $R\simeq 10^{12}\,\mathrm{cm}$, and  luminosity of the X-ray source $L\simeq 10^{38}\,\mathrm{erg}\,\mathrm{s}^{-1}$.

We have also made a very simple simulation  to estimate the range of ionization parameters which would be expected in \cyg. %
We assume that the gas is supplied by the stellar wind, which is spherically symmetric around the star.  The total wind mass loss rate is $10^{-5}\,M_\odot$ yr$^{-1}$ and the terminal velocity is 1000\,\kms, with a `beta law', i.e. velocity vs. radius is given by $v(R)\simeq v_\infty(1-R_*/R)^\beta$ with $\beta=1$.   The X-ray source is assumed to be isotropic with luminosity $10^{38}\,\mathrm{erg}\,\mathrm{s}^{-1}$. We find that the average ionization parameter in the wind is $\log\xi{\sim}3$. This is approximately constant over a large region since both the wind and radiation density are diluted at the same rate at large distances.   The ionization parameter can be much greater close to the X-ray source,  $\log\xi\sim$4--5, and much smaller close to the star.   Although the dynamics and density distribution in this simulation are highly idealized, this demonstrates that the wind can supply gas at approximately the right ionization parameter  to account for the absorption component that we observe in \cyg.

\section{Conclusions}

The \rsl spectrum of \cyg shows exceptionally strong emission and absorption in lines from iron, and from other abundant elements.   These lines come from partially ionized species though there is evidence for emission from near-neutral iron. %
The iron K$\alpha$ lines show a complex interlocking structure in which resonance absorption partially suppresses emission and shifts the apparent peaks of the emission lines.  We fit to photoionization models for the emission and absorption components, derive Doppler shifts, line broadening and plasma conditions. We show that the lines come from a range of ionic species broader than is typical for single-zone photoionization models.  The absorption line shifts are generally consistent with a single blueshift, corresponding to a speed $-534\pm6\,\kms$.  The average velocity shift of the emission lines is significantly smaller. %
We use the orbital phase variability to infer possible locations and kinematics for the various absorption and emission components.    We show that the absorption line Doppler shift varies with a semi-amplitude of $55\pm7\,\kms$ around the orbit, while the emission line Doppler shift varies by $194\pm 29\,\kms$.  This suggests that the absorption comes from the wind or associated structures on a size scale comparable to the binary orbit.  The strengths of the absorption lines are crudely consistent with what is known about the primary wind.  The orbital phase dependence of the emission is consistent with being fixed relative to the compact object in its orbit. %
The luminosity of the emission lines is consistent with emission from gas which is photoionized by the observed X-ray continuum.

\begin{acknowledgements}

We thank the anonymous referee for an insightful and constructive review that improved this work.  Part of this work was performed under the auspices of the U.S. Department of Energy by Lawrence Livermore National Laboratory under Contract DE-AC52-07NA27344. The material is based upon work supported by NASA under award number 80GSFC21M0002. This work was supported by JSPS KAKENHI grant numbers JP22H00158, JP22H01268, JP22K03624, JP23H04899, JP21K13963, JP24K00638, JP24K17105, JP21K13958, JP21H01095, JP23K20850, JP24H00253, JP21K03615, JP24K00677, JP20K14491, JP23H00151, JP19K21884, JP20H01947, JP20KK0071, JP23K20239, JP24K00672, JP24K17104, JP24K17093, JP20K04009, JP21H04493, JP20H01946, JP23K13154, JP19K14762, JP20H05857, and JP23K03459. This work was supported by NASA grant numbers 80NSSC20K0733, 80NSSC18K0978, 80NSSC20K0883, 80NSSC20K0737, 80NSSC24K0678, 80NSSC18K1684, and 80NNSC22K1922. LC acknowledges support from NSF award 2205918. CD acknowledges support from STFC through grant ST/T000244/1. LG acknowledges financial support from Canadian Space Agency grant 18XARMSTMA. MS acknowledges the support by the RIKEN Pioneering Project Evolution of Matter in the Universe (r-EMU) and Rikkyo University Special Fund for Research (Rikkyo SFR). AT and the present research are in part supported by the Kagoshima University postdoctoral research program (KU-DREAM). SY acknowledges support by the RIKEN SPDR Program. IZ acknowledges partial support from the Alfred P. Sloan Foundation through the Sloan Research Fellowship. KIIK has received funding from the European Research Council (ERC) under the European Union’s Horizon 2020 research and innovation program (grant agreement No. 101002352, PI: M. Linares). This work was supported by the JSPS Core-to-Core Program, JPJSCCA20220002. The material is based on work supported by the Strategic Research Center of Saitama University.

\end{acknowledgements}

\bibliographystyle{aasjournal} \bibliography{references}

\end{document}